\documentclass[apjl]{emulateapj}

\slugcomment{Accepted for Publication to  ApJL}

\shorttitle{Accurate Low-Mass Stellar Models of KOI-126}
\shortauthors{Feiden et al.}

\begin{document}

\title{Accurate Low-Mass Stellar Models of KOI-126}

\author{Gregory A. Feiden\altaffilmark{1}, Brian Chaboyer\altaffilmark{1}, and Aaron Dotter\altaffilmark{2}}

\altaffiltext{1}{Department of Physics and Astronomy, Dartmouth College, 6127 Wilder Laboratory, Hanover, NH 03755}
\altaffiltext{2}{Space Telescope Science Institute, 3700 San Martin Dr., Baltimore, MD 21218}

\begin{abstract}
The recent discovery of an eclipsing hierarchical triple system with two low-mass stars in a close orbit
(KOI-126) by Carter et al. (2011) appeared to reinforce the evidence that theoretical stellar evolution 
models are not able to reproduce the observational mass-radius relation for low-mass stars. We present
a set of stellar models for the three stars in the KOI-126 system that show excellent agreement with the observed
radii. This agreement appears to be due to the equation of state implemented by our code.
A significant dispersion in the observed mass-radius relation for fully convective stars is demonstrated; 
indicative of 
the influence of physics currently not incorporated in standard stellar evolution models. We also 
predict apsidal motion constants for the two M-dwarf companions. These values should be 
observationally determined to within 1\% by the end of the \emph{Kepler} mission. 
\end{abstract}


\section{Introduction}

Double-lined eclipsing binaries (hereafter DEBs) are powerful tools for testing stellar evolution
models. A wealth of information can be gleaned from observations of DEBs, including precise masses 
and radii of the component stars along with the apsidal motion of the system, if the orbit is sufficiently
eccentric \citep[see][for a review]{Torres2010}. To date, observations of DEB systems with at least one low-mass
component (below 0.8 $M_{\odot}$) have painted a grim picture for stellar evolution models. The radii 
predicted by models are systematically 5~-~10\% smaller than those determined from observations, while the
effective temperatures derived from models are 5\% cooler \citep[e.g.,][and references therein]{Ribas2008,Morales2008,Morales2009,Torres2010}.

The aforementioned discrepancies have been attributed to the effects of large scale magnetic fields \citep{Ribas2006,Chabrier2007},
as most of the low-mass DEBs discovered are close binary systems with short orbital periods
(generally less than 3 days). Tidal interactions act to spin-up the stars as they progress towards complete tidal
synchronization. Short rotation periods result from this process, amplifying the stellar magnetic 
field strength. Strong magnetic fields inhibit the efficiency of convective energy transport
and increase the total surface coverage of starspots, effectively lowering the temperature at the 
stellar surface. In order to conserve flux, the stellar radius is forced to inflate \citep[e.g.,][]{
Gough1966,Mullan2001,Chabrier2007}.

Recently, \citet[henceforth C11]{Carter2011} discovered a triply eclipsing hierarchical
triple system, \object{KOI-126}, in the \emph{Kepler} data set which contains two low-mass stars (KOI-126 B, C).
The two low-mass stars are in a tight 1.77 day orbit itself orbiting a more massive primary star (KOI-126 A) on a fairly eccentric 
path about every 34 days. The authors were able to derive fundamental parameters for all three stars:
$M_A =$~1.347~$\pm$~0.032~$M_{\odot}$, $R_A =$~2.0254~$\pm$~0.0098~$R_{\odot}$,
$M_B =$~0.2413~$\pm$~0.0030~$M_{\odot}$, $R_B =$~0.2543~$\pm$~0.0014~$R_{\odot}$, 
$M_C =$~0.2127~$\pm$~0.0026~$M_{\odot}$, and $R_C =$~0.2318~$\pm$~0.0013 $R_{\odot}$.
Spectroscopy of the more massive primary indicated it has an effective temperature of 5875~$\pm$~100~K with 
a super-solar metallicity ([Fe/H]~=~+0.15~$\pm$~0.08). Lastly, the relatively large eccentricity of the
systems allowed the authors to place a weak constraint on the apsidal motion constant for the two low-mass 
stars, determining that it is below 0.6 at the 95\% confidence level.

C11 derived an age for the system of 4 $\pm$ 1 Gyr after fitting KOI-126~A to theoretical Y$^{2}$ isochrones \citep{Demarque2004}.
When C11 compared the observed mass-radius relation to 1, 2, 4, and 5 Gyr \citet[BCAH98]{Baraffe1998}
isochrones for [Fe/H] $\le$ 0.0, the models were seen to under-predict the radius of the two low-mass 
stars by 2-5\%. However, it is important to note that no super-solar metallicity isochrones are 
available for the BCAH98 models (in part due to limitations of their equation of state). C11
suggested that the observed radius discrepancy could be the result of both the super-solar metallicity
and possible magnetic activity of the system. 

Here, we present the results of theoretical stellar modeling of the KOI-126 
components using the Dartmouth Stellar Evolution Program \citep[DSEP]{Chaboyer2001,Bjork2006,Dotter2008}, a 
descendant of the Yale Rotating Stellar Evolution Code \citep{Guenther1992}. We derive model
radii consistent with observations as well as theoretical apsidal motion constants for the two low-mass
stars. In \S 2 we describe the DSEP models utilized for this study while in \S 3 we present our 
primary results. Finally, in \S 4, we discuss the implications of this study with regards to the only other well known
low-mass eclipsing system, CM Dra.

\section{Models}
We constructed individual stellar evolution models for each KOI-126 component along with a series of 
theoretical isochrones using DSEP\footnote{http://stellar.dartmouth.edu/$\sim$models/}. The physics 
incorporated in the models have been described previously \citep{Chaboyer2001,Bjork2006,Dotter2007,Dotter2008}, but we shall
provide a brief summary. 

Our models include the effects of helium and heavy element diffusion following
the prescription of \citet{Thoul1994}, though for fully convective stars, diffusion physics are unimportant.
The opacities utilized by DSEP are the high-temperature OPAL opacities \citep{Iglesias1996}
and low-temperature opacities of \citet{Ferguson2005}. Surface boundary conditions were defined using the PHOENIX
model atmospheres \citep{Hauschildt1999a,Hauschildt1999b} and were attached to the interior model at $T=T_{eff}$ by
interpolating in model atmosphere tables. Attaching the atmospheres to $T=T_{eff}$ makes the value of the convective 
mixing length used in the atmosphere code inconsequential \citep{BCAH97}.

Above 0.8 $M_{\odot}$, DSEP uses a general ideal gas equation of state 
with the Debye-H\"{u}ckel correction \citep{Chaboyer1995}. In the low-mass regime, DSEP
employs the FreeEOS\footnote{By Alan Irwin: http://freeeos.sourceforge.net} in the EOS4 configuration, selected for
its treatment of arbitrary heavy element abundances and its inclusion of the H$_{2}^+$ molecule. FreeEOS has also
been shown to be valid for modeling stars more massive than 0.1 $M_{\odot}$ \citep{Irwin2007}. Convective 
core overshoot is included using the method of \citet{Demarque2004}. Rotation was not considered.

The only modification made to the underlying physics in DSEP is related to the partial inhibition of element diffusion. 
We have introduced turbulent diffusion as described by \citet{Richard2005}. Turbulent
diffusion modifies the atomic diffusion coefficient and acts to extend the mixing region below the convection zone.
The magnitude of the turbulent diffusion coefficient is tied to an adjustable reference temperature, $T_{0}$, and varies 
with density via:
\begin{equation}
D_{T} = \omega D_{He}\left(T_{0}\right)\left[\frac{\rho}{\rho(T_{0})}\right]^{-3}
\end{equation}
where $\omega$ characterizes the relative strength of turbulent diffusion and $D_{He}(T_{0})$ and $\rho(T_{0})$ are the helium 
diffusion coefficient and density at the prescribed reference temperature, respectively. \citet{Proffitt1991} motivate the 
$\rho^{-3}$ dependence in order to reproduce the solar beryllium abundance, which appears to be unchanged over time. Thus, 
any non-standard mixing in the Sun must be localized to a narrow region below the solar convection zone. We select 
$\omega = 400$ and leave it fixed, in concordance with \citet{Richard2005}. The reference temperature used primarily in
our models is $T_{0} = 10^{6}$, which was found to best reproduce the observed abundance trends of \object{NGC 6397} \citep{Korn2007}.

A solar calibration model was generated to determine the appropriate
initial mass fractions of helium ($Y_{init}$) and metals ($Z_{init}$), given the solar heavy element composition of 
\citet{GS98}, as well as to calibrate the convective mixing length, $\left(\alpha_{MLT} = \ell / H_P\right)$. At the
solar age \citep[4.57 Gyr;][]{Bahcall2005} we were able to reproduce the solar radius, solar luminosity, radius of the 
convective boundary, and $(Z/X)_{\odot}$ with $\alpha_{MLT}$~=~1.938, $Y_{init}$~=~0.27491, and $Z_{init}$~=~0.01884.
All of the models utilized in this study were calculated using the solar calibration as a reference and were assumed
to be coeval. Super-solar metallicity models with [Fe/H] = +0.15 were generated with $Y_{init}$~=~0.28419 
and $Z_{init}$~=~0.02469.

\section{Results}
\subsection{Stellar Age}
The age of the system was constrained by evolving a 1.347~$M_{\odot}$ model, with [Fe/H]~=~+0.15 and 
solar calibrated $\alpha_{MLT}$, and matching the model radius with the observed radius of KOI-126 A (Figure 1). 
The age we derive for the system is 4.1 $\pm$ 0.6 Gyr, consistent with the age derived by C11. Our uncertainty 
in the age is dominated by the uncertainty in the observed mass and metallicity of the system, with the observational
uncertainty of the radius being of negligible importance. The effect of changing the reference temperature for turbulent
diffusion was also considered (including removing it entirely) and was found to play a negligible role in the age 
determination.

\begin{figure}
\plotone{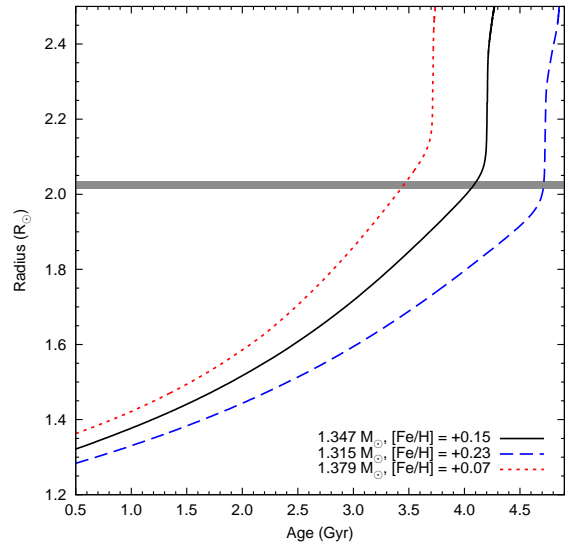}
\caption{Stellar evolution tracks used to constrain the age of KOI-126~A. The dark band signifies the radius constraints
imposed by the observations.}
\end{figure}

\subsection{Mass-Radius Relation}
The primary results of this Letter are demonstrated in Figure 2. DSEP accurately reproduces the observed 
radius for each of the low-mass stars at 4.1 Gyr with [Fe/H]~=~+0.15. We find that for masses 
M$_2$~=~0.2410~$M_{\odot}$ and M$_3$~=~0.2130~$M_{\odot}$ the predicted radii from DSEP are R$_2$~=~0.2544~$R_{\odot}$ and 
R$_3$~=~0.2312~$R_{\odot}$, indicating a relative error between the model and observed radii of less than 0.3\%. 
At solar metallicity, the models predict radii approximately 1\% smaller than those predicted by the super-solar
metallicity models. The predicted radii are robust. Artificially reducing the mixing length ($\alpha_{MLT}=1.00$)
and fitting the atmosphere to a deeper point in the stellar envelope ($\tau =100$) both 
produced radius changes under 0.5\%.

Solar metallicity isochrones from DSEP display radii approximately 1\% larger than radii predicted by BCAH98.
The difference is likely a consequence of the equation of state (EOS)
utilized by each group. Whereas the EOS used by BCAH98 is calculated for a pure hydrogen/helium plasma, FreeEOS 
calculates the EOS for an arbitrary metal abundance. Our models can be more reliably calculated above solar
metallicity. To test the effect of the EOS, we ran DSEP with the \citet[][SCVH]{scvh95} EOS as an attempt to mimic
the BCAH98 model radii predictions. Our models using the SCVH EOS produced radii within 0.5\% of the BCAH98 models
at the same mass and composition, illustrating the importance of the EOS.

\begin{figure}
\plotone{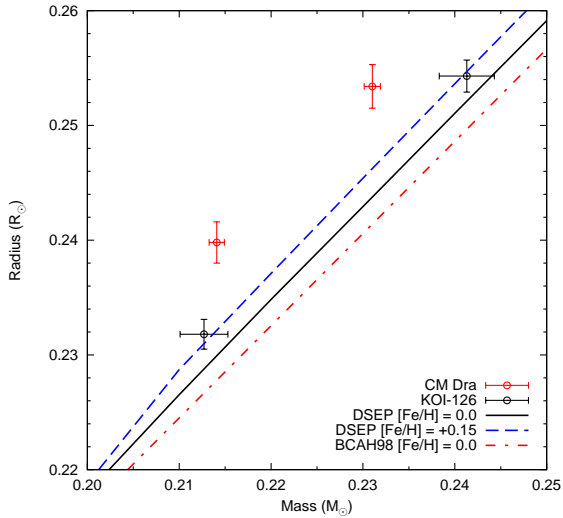}
\caption{Mass-radius relationship as defined by CM Dra (red -- points) and KOI-126 (black -- points). Overlaid
are 4.1 Gyr theoretical isochrones from DSEP with [Fe/H] = 0.0 (black -- solid) and [Fe/H] = 0.15 (blue -- dash) and
from BCAH98 with [Fe/H] = 0.0 (red -- dash-dot).}
\end{figure}

\subsection{Relative Fluxes}
The effective temperature of a star is another measure with which to compare our models. While C11 found the
effective temperature of KOI-126~A to be 5,875~$\pm$~100~K from spectroscopy (which our models match), they were not able to determine
the effective temperatures of the two M-dwarfs. However, the dynamical/photometric model utilized by C11 yielded
the flux for each M-dwarf relative to the primary. The model results were: $f_{B}/f_{A} = (3.26\pm0.24)\times 
10^{-4}$ and $f_{C}/f_{A} = (2.24\pm0.48)\times 10^{-4}$. 

To compare our stellar models, we derived a synthetic color-$T_{eff}$ transformation for the \emph{Kepler} 
bandpass using the spectral response function provided in the \emph{Kepler Instrument Handbook} \citep{vancleve2009}.
With model fluxes determined using the PHOENIX model atmospheres, it was possible to generate synthetic colors for each
of the model components of KOI-126. We found $f_{B}/f_{A} = 4.77\times 10^{-4}$ and $f_{C}/f_{A} = 3.71\times 10^{-4}$,
a 6$\sigma$ and 3$\sigma$ difference, respectively. We also derived
relative fluxes utilizing the empirical relation given on the \emph{Kepler} Guest Observer website\footnote{See http://keplergo.arc.nasa.gov/CalibrationZeropoint.shtml}.
The empirical relation relies on the SDSS \emph{g} and \emph{r} magnitudes to derive the approximate 
\emph{Kepler} magnitude. SDSS magnitudes for our models were calculated using a synthetic color-$T_{eff}$ 
transformation \citep{Dotter2008}, allowing for an estimate of the \emph{Kepler} magnitudes. This semi-empirical
transformation yielded: $f_{B}/f_{A} = 3.07\times 10^{-4}$ and $f_{C}/f_{A} = 2.36\times 10^{-4}$, consistent with
the relative fluxes derived by C11.

In an effort to decrease the relative flux for each M-dwarf in the purely theoretical transformation, we varied the 
parameters of the primary star within
the given observational constraints and changed the amount of convective core overshoot (CCO). When considering
the largest mass (1.379 M$_{\odot}$), lowest metallicity (+0.07), and a relatively high amount of CCO, the relative
fluxes were reduced to within 2$\sigma$ of the accepted values. As with the age determination, the tight observational
constraint imposed on the radius of the primary made the range of relevant radii insignificant in deriving our results.

\subsection{Apsidal Motion Constant}

\begin{deluxetable*}{c c c c c c c c c}
\tabletypesize{\scriptsize}
\tablecaption{Apsidal motion constants (k$_2$) for fully convective stars with 
[Fe/H] = -0.50, 0.0, and +0.15. Values are quoted at 7 different ages (in Gyr).
\label{apcon}}
\tablewidth{0pt}
\tablehead{
\colhead{Mass ($M_{\odot}$)} & \colhead{[Fe/H]} & \colhead{0.5} & \colhead{1.0} &
\colhead{2.0} & \colhead{3.0} & \colhead{4.0} & \colhead{5.0} & \colhead{10.0}
}
\startdata
0.20 & -0.50 & 0.1523 & 0.1523 & 0.1522 & 0.1522 & 0.1522 & 0.1522 & 0.1521 \\
\nodata & 0.00 & 0.1521 & 0.1521 & 0.1521 & 0.1521 & 0.1521 & 0.1520 & 0.1519 \\
\nodata & +0.15 & 0.1518 & 0.1517 & 0.1517 & 0.1517 & 0.1517 & 0.1517 & 0.1516 \\
0.25 & -0.50 & 0.1499 & 0.1498 & 0.1498 & 0.1498 & 0.1498 & 0.1497 & 0.1496 \\
\nodata & 0.00 & 0.1496 & 0.1496 & 0.1496 & 0.1496 & 0.1495 & 0.1495 & 0.1494 \\
\nodata & +0.15 & 0.1496 & 0.1496 & 0.1496 & 0.1495 & 0.1495 & 0.1495 & 0.1493 \\
0.30 & -0.50 & 0.1482 & 0.1482 & 0.1481 & 0.1481 & 0.1481 & 0.1480 & 0.1479 \\ 
\nodata & 0.00 & 0.1480 & 0.1480 & 0.1479 & 0.1479 & 0.1478 & 0.1478 & 0.1477 \\
\nodata & +0.15 & 0.1479 & 0.1479 & 0.1479 & 0.1478 & 0.1478 & 0.1477 & 0.1476 \\
\enddata
\end{deluxetable*}

By the end of the nominal \emph{Kepler} mission, C11 predict they will know
the apsidal motion constant of the two low-mass stars with a relative precision
of nearly 1\%. From our low-mass stellar models, we were able to predict an apsidal motion 
constant for each star, quantifying the degree to which the mass within the star is centrally 
concentrated. To do this we followed the prescription of \citet{Kopal1978} and solved Radau's equation
with $j = 2$:
\begin{equation}
r \frac{d\eta_{2}}{dr} + \frac{6\rho (r)}{\left<\rho\right>}\left(\eta_{2} + 1\right) + \eta_{2}\left(\eta_{2} - 1\right) = 6
\end{equation}
where $\eta_{2}(r)$ is the value of a particular solution to Radau's equation related to
the stellar deviation from sphericity,
$\rho(r)$ is the density of the plasma, and $\left<\rho\right>$ is the average 
density. We used a $4^{th}$ order Runge-Kutta integration method to solve Radau's equation 
to obtain a particular solution at the surface of each star. Having 
determined $\eta_2 (R)$, we were able to determine the apsidal motion constant via:
\begin{equation}
k_{2} = \frac{3 - \eta_{2}(R)}{4 + 2\eta_{2}(R)}
\end{equation}
As a check on our apsidal motion constant integrator, we generated polytropic models characterized by
the polytropic constant n = 1.0 and 1.5 and compared the results of our code with the results of 
\citet{Brooker1955}.

The apsidal motion constants
derived from the interior structure of our models for KOI-126 B and C are $k_2$~=~0.1499 and $k_2$~=~0.1512, 
respectively. This suggests our models are slightly less centrally condensed than an n = 1.5 polytrope which
is characterized by $k_{2}$~=~0.1433. In fact, the run of density and pressure for our models indicate our 
models are best described by a polytrope with n~$\sim$~1.45. A set of theoretical apsidal motion constants
for fully convective stars, generated using DSEP, is given in Table~\ref{apcon} and may be compared to future
observations. Finally, we note the effects of rotation on the derived $k_2$ values are negligible.
Employing the formulae of \citet{Stothers1974}, we find that rotation affects our $k_2$ values at the 0.02\%
level for stars rotating with a period of 1.7 days.

\section{Discussion}
Low-mass stars below the fully convective boundary are a wonderful tool to test basic physics. 
Their low mass affords theorists a stable, long lived ($> 10^{11}$ yr) laboratory with
which to test the physics incorporated in the models. These fully convective stars have 
relatively simple structures, and uncertainties in the opacities, surface boundary conditions, 
and the treatment of convection have relatively small effects on the predicted properties of 
the models \citep[ \S~3.2]{DotterTh}. Modelers are relieved of having to specify a free parameter
since the internal structure is insensitive to the prescribed mixing length. The age of these stars
is also of little consequence, as the radii of low-mass stars is nearly constant over
their main sequence lifetime. These effects imply a unique mass-radius relation for stars of a 
given composition below the fully convective boundary. The discovery of two new data points in this 
regime indicate the mass-radius relation is not unique and there is significant dispersion amongst the data. 

Discrepancies are still seen between our models and the
components of \object{CM Draconis} \citep[hereafter CM Dra]{Morales2009}, as shown in Figure 2. Comparing 
KOI-126 and CM Dra, both have low-mass components with strikingly similar masses, estimated ages, 
and orbital periods (1.77~d and 1.27~d, respectively). Both systems are thought 
to be tidally synchronized, though not circularized. Their similar characteristics suggest the
M-dwarfs in both systems should have similar radii. Not only do they not,
but the opposite relationship of what is theoretically expected is observed. CM Dra has a sub-solar
metallicity \citep[][and references therein]{Morales2009} and should therefore possess smaller radii than KOI-126.

It is possible that the KOI-126 system has been caught in a period of inactivity, similar 
to a solar minimum. CM Dra appears to have undergone such a period of quiescence in 2000.
Morales et al. noted that no corrections due to starspots were needed in the analysis of the light curve 
data from that year. If KOI-126 B and C are in a magnetically quiescent state, it is possible 
that their radii would not show signs of starspots or inflated radii. We note that starspots
would likely be of negligible importance since variations due to spots would be reduced to noise amongst the
signal of KOI-126~A.

We find it encouraging that DSEP is able to predict the mass-radius relation suggested by KOI-126 B and C.
Although, more data points are needed to allow for a more complete understanding of the reliance of the
mass-radius relation on physics (standard and non-standard) incorporated in current models.
It is clear from this work that no standard stellar evolution model will be
able to simultaneously fit both CM Dra and KOI-126 and that work on non-standard stellar evolution models
will be required to fit CM Dra. As suggested by numerous authors, magnetic activity 
is likely the culprit and must be incorporated into the next generation of models. However, the effects of a magnetic field on the 
interior structure of stars has previously been considered as a reduction in the prescribed mixing length, 
mimicking the reduction in convective efficiency that should accompany the presence of a magnetic field. With
this in mind, it is not clear how magnetic activity would affect stars below the fully convective boundary.
Self-consistent magnetic stellar models should help lend insight into the discrepancies
with CM Dra.

Concerning the relative flux discrepancies observed between the purely theoretical transformation and the 
photometric models of C11, the color-$T_{eff}$ transformations are of interest and might not provide
a fully accurate transformation to the observational plane. In the low-mass regime, opacities 
are complicated by the formation of molecules and the peak of the stellar spectrum is near the cut-off of 
the \emph{Kepler} response function. Interestingly, the semi-empirical transformation yields
relative fluxes which are entirely consistent with the photometric models of C11. However, we must be cautious
with this result as the systematic errors are not well constrained for the transformation from the SDSS magnitudes
to the \emph{Kepler} magnitude. It must also be noted that the
flux of the primary star is sensitive to the details of CCO. We observed that increasing the amount of CCO brought
the purely theoretical fluxes closer to the observational values. More investigation will be required to accurately diagnose the
discrepancies.

Finally, we note that the determination of the apsidal constant will provide a crucial test of our stellar
evolution models. In particular, it will test the EOS, which directly determines the run of density necessary for the computation
of the apsidal motion constant. Morales et al. found that using the BCAH98 models, $k_2$ was approximately 0.11
whereas our models predict a larger value of approximately 0.15. The difference between these two values is directly 
attributable to the EOS, which determines the run of density within a stellar model. If C11 are able to accurately derive the
apsidal motion constant to within 1\%, it will provide a stringent benchmark against which to
test the interior physics of low-mass stellar evolution models.

\section{Summary}
In their discovery paper, C11 reported that the triply eclipsing hierarchical triple KOI-126 appeared to
support the mounting evidence that current standard low-mass stellar models are unable to reproduce
the observed mass-radius relation. However, we have generated stellar models and
theoretical isochrone tracks using the Dartmouth Stellar Evolution Program and find our model
radii agree with the observations. Combining the KOI-126 measurements with previous observations of
the low-mass binary system CM Dra, we find that the dispersion in the observed fully convective mass-radius
relation is significant and stands in contrast to theoretical predictions. The fact that CM Dra, a system
with sub-solar metallicity, lies on the super-solar side of the theoretical mass-radius relation is indicative
of physics currently not incorporated in standard stellar models. We predict the apsidal motion
constant for each of the KOI-126 low-mass stars and find $k_{2} \simeq$~0.15. C11 postulate 
that they will be able to determine the apsidal motion constant with a relative precision of 1\% by the 
end of the nominal \emph{Kepler} mission. This will provide a crucial test for our models and will provide a stringent 
constraint against which to test all current and future low-mass standard stellar evolution models.

\acknowledgments

We are grateful for the National Science Foundation (NSF) grant AST-0908345, which supported this work.
Gratitude is also given to the anonymous referees whose comments and suggestions have improved the 
quality of this manuscript. We wish to thank F. Allard for illuminating some of the particulars of 
the PHOENIX atmosphere code.

\end{document}